\documentclass[aps,prl,twocolumn,nofootinbib,longbibliography,10pt]{revtex4-2}
\usepackage{etoolbox}
\usepackage{dcolumn}
\usepackage{amsmath,amssymb,amsfonts,mathtools}
\usepackage{mathrsfs,bbold}
\usepackage{graphicx}
\usepackage[colorlinks=true,urlcolor=blue,citecolor=red,linkcolor=blue]{hyperref}
\usepackage{accents}
\newlength{\dhatheight}
\usepackage{natbib}
\usepackage{wrapfig}
\usepackage{flushend,BOONDOX-cal,BOONDOX-frak}

\makeatletter
\newsavebox{\@brx}
\newcommand{\llangle}[1][]{\savebox{\@brx}{\(\m@th{#1\langle}\)}%
  \mathopen{\copy\@brx\kern-0.5\wd\@brx\usebox{\@brx}}}
\newcommand{\rrangle}[1][]{\savebox{\@brx}{\(\m@th{#1\rangle}\)}%
  \mathclose{\copy\@brx\kern-0.5\wd\@brx\usebox{\@brx}}}
\makeatother

\begin{document}
\title{{Bose-Einstein condensate as a quantum gravity probe}\\{``\textit{Erste Abhandlung}"}}
\author{Soham Sen}
\email{sensohomhary@gmail.com}
\affiliation{Department of Astrophysics and High Energy Physics, S. N. Bose National Centre for Basic Sciences, JD Block, Sector-III, Salt Lake City, Kolkata-700 106, India}
\author{Sunandan Gangopadhyay}
\email{sunandan.gangopadhyay@gmail.com}
\affiliation{Department of Astrophysics and High Energy Physics, S. N. Bose National Centre for Basic Sciences, JD Block, Sector-III, Salt Lake City, Kolkata-700 106, India}
\begin{abstract}
\noindent We consider a Bose-Einstein condensate interacting with a gravitational wave for the case when the gravitational fluctuations are quantized in order to incorporate quantum gravity effects into the theory. We observe that the solution of the time-dependent part of the pseudo-Goldstone boson has infusions from the noise induced by gravitons and the corresponding differential equation of motion is Langevin-like. Using this result, we obtain the quantum gravity modified Fisher information which has been termed as the quantum gravitational Fisher information (QGFI). The inverse square root of the stochastic average of the QGFI gives the minimum uncertainty in the measurement of the gravitational wave amplitude. The minimum uncertainty does not go to infinity as the measurement time approaches zero in a quantum gravity setup rather it has a measurable finite value for gravitons with high squeezing. Finally, we observe the effect of decoherence due to interacting phonon modes in the QGFI and observe a less obvious decoherence effect for higher squeezing of the initial graviton.
\end{abstract}
\maketitle
\section{Introduction}\label{S1}
\noindent Detecting the quantum nature of gravity has been one of the most fundamental pursuits of theoretical high energy physics in recent times. It has already been observed in several failed theoretical attempts that the full theory of gravity is non-renormalizable. The only silverlining of hope is that the linearized quantum gravity theory is perfectly renormalizable and the spin-2 mediators of gravitational interaction are termed as gravitons. To directly detect a graviton is an impossible task, hence, people have rather tried to investigate signatures of quantum gravity via the consideration of quantum gravitational interaction in very well known physical (semi-classical as well as quantum mechanical) scenarios. With the detection of gravitational waves \cite{GWaveDetection,GWaveDetection2,
GWaveDetection3}, the primary target of seeing signatures of gravitons has been the upcoming new gravitational wave observatories (LISA, DECIGO etc.) which are capable of working at much higher sensitivities than the ones currently under use (LIGO and VIRGO). On any classical equation of motion it has been observed in several literatures \cite{QGravNoise,QGravD,QGravLett,KannoSodaTokuda,KannoSodaTokuda2,AppleParikh,OTMGraviton} that the gravitons induce a stochastic effect resulting in a Langevin-like equation of motion. Solving this equation of motion, one can arrive at a soluton which leads to a standard deviation in the measurement of the position variable. This standard deviation scales exponentially with the squeezing of the incoming gravitons and with a high enough squeezing it may be possible to detect the signature of linearized quantum gravity in future generation of gravitational wave detectors. Recently we have also obtained the uncertainty relation for a graviton-matter interaction case \cite{OTMApple} and observed that it is dependent on the square of the variance in the momentum parameter.

\noindent Another important aspect from the low-temperature physics is a Bose-Einstein condensate which came as a consequence of the joint efforts by Satyendranath Bose \cite{SNBose} and Albert Einstein \cite{Einstein1,Einstein2,Einstein3}. The reports of the first experimental construction of  a Bose-Einstein condensate (BEC) came out in the year 1995 \cite{1Nobel2001,2Nobel2001}. Very recently there has been a novel work involving the detection of gravitational waves using a Bose-Einstein condensate \cite{PhononBEC}. In this work, a quasi (1+1)-dimensional BEC with fluctuating boundary condition has been analyzed. Another analysis was done in \cite{PhononBEC2} using a non relativistic BEC-gravitational wave interaction model. Finally in \cite{PhononBEC3,PhononBEC4}, a (3+1)-dimensional generalization of the model presented in  \cite{PhononBEC} was done. The quantum metrological techniques developed in \cite{BraunsteinCaves,Uhlmann2mode,quantum_metrology}, has been used here to obtain the quantum Fisher information. The quantum Fisher information then leads one to estimate the standard deviation in the amplitude parameter of the gravitational wave and its lower bound \cite{PhononBEC3,PhononBEC4,ThesisMatthew}
.\\																			\noindent Our primary aim is to use the zero-temperature Bose-Einstein condensate to detect signatures of quantum gravity. At first we have used the self-interacting bosonic complex scalar field theory in curved spacetime combined with the Einstein-Hilbert action depicting the dynamics of the gravitational fluctuation term. Using the principle of least action, one can then obtain the equation of motion corresponding to the variables in concern. Next we quantize the theory by raising the phase space variables to operator status and imposing suitable canonical commutation relations among the conjugate variables. The equation of motion corresponding to the time dependent part of the pseudo-Goldstone boson now obeys a Langevin-like stochastic differential equation. The noise term has a vanishing one point correlator and a non-vanishing two-point correlator. Using the methods of quantum metrology, we have then obtained the quantum gravity induced Fisher information utilizing the perturbative solution of the stochastic differential equation. In our case, the Fisher information becomes stochastic in nature because of the noise induced by the gravitons. This quantum gravity infused Fisher information has been termed as a quantum gravitational Fisher information (QGFI). The QGFI is then used to obtain the optimum lower bound to the standard deviation in the gravitational wave amplitude. We have then calculated the standard deviation in the Fisher information as well.	We have also calculated the lower bound to the standard deviation in the amplitude parameter for all modes of the BEC when the graviton-noise harbours a Gaussian decay term. Finally, we have considered a more realistic scenario where there is decoherence in the system due to interaction between the phonon modes and we have observed the effect of squeezing of the gravitons on the overall dissipation of the system.																																	
\section{Model system and the noise induced by gravitons} \label{S2}
\noindent In this section, we shall first discuss the background of the model and discuss the effects of the noise induced by gravitons. It is important to note that we are working in a mostly positive signature. The Einstein-Hilbert action in the transverse-traceless gauge condition takes the form
\begin{equation}\label{L1}
S_{EH}=-\frac{1}{8\kappa^2}\int d^4 x~ \partial_{\kappa}\bar{h}_{ij}\partial^{\kappa}\bar{h}^{ij}
\end{equation}
where $\kappa=\sqrt{8\pi G}$. Using the Fourier mode decomposition of the fluctuation term $\bar{h}_{ij}$ inside a finite size box of volume $V$ as $\bar{h}_{ij}(t,\mathbf{x})=\frac{2\kappa}{\sqrt{V}}\sum_{\mathbf{k},s}h^s(t,\mathbf{k})e^{i\mathbf{k}\cdot\mathbf{x}}\epsilon^s_{ij}(\mathbf{k})$, one can obtain the new form of the Einstein-Hilbert action as
\begin{equation}\label{L2}
\begin{split}
S_{\text{EH}}=\frac{1}{2}\sum_{\textbf{k},s}\int dt\left(\bigr|\dot{h}^s(t,\mathbf{k})\bigr|^2-k^2\bigr|h^s(t,\mathbf{k})\bigr|^2\right)~.
\end{split}
\end{equation}
If the complex scalar field $\phi(t,\mathbf{x})$ corresponds to a homogenous bosonic field, then one can express it in terms of the real heavy field $\varphi(t,\mathbf{x})$ and $\chi(t,\mathbf{x})$ as $\phi(t,\mathbf{x})=e^{i\chi(t,\mathbf{x})}\varphi(t,\mathbf{x})$, where the self interaction term of the Lagrangian is given by $\lambda|\phi(t,\mathbf{x})|^4$. Extremizing the Lagrangian of the theory with respect to the heavy fields and integrating out all of the higher order derivative contributions from the theory \cite{PhononBEC3,DTSon}, one arrives at the final form of the action corresponding to the bosonic part of the system as
\begin{equation}\label{L3}
S_{\text{BEC}}=\int d^4x\sqrt{-g}\mathcal{L}_{\text{BEC}}
\end{equation}
where the Lagrangian of the Bose-Einstein condensate (BEC) reads
\begin{equation}\label{L4}
\mathcal{L}_{\text{BEC}}=\frac{1}{4\lambda}\left(g_{\mu\nu}\partial^\mu\chi\partial^\nu\chi+m^2\right)^2.
\end{equation}
If $\pi(t,\mathbf{x})\in\mathbb{R}$ denotes the phonons corresponding to the Bose-Einstein condensate (also called the pseudo-Goldstones bosons), we can express $\chi(t,\mathbf{x})$ as
$\chi(t,\mathbf{x})=\tilde{\sigma}x_\mu\delta^{\mu}_{~0}+\pi(t,\mathbf{x})$.
One can express the pseudo-Goldstone bosons as $\pi(t,\mathbf{x})=\sum_{\mathbf{k}_\beta}e^{i\mathbf{k}_\beta\cdot\mathbf{x}}\psi_{\mathbf{k}_\beta}$. Considering a box of volume $V_\beta$ with sides $L_\beta$ for the quantization of the Goldstone bosons, and using the BEC dispersion relation $\omega_\beta\simeq c_sk_\beta$ \cite{PhononBEC3,PhononBEC4} $\left(\text{with } c_s^2=\frac{\tilde{\sigma}^2-m^2}{3\tilde{\sigma}^2-m^2}\right)$, we can write down the BEC Lagrangian as
\begin{equation}\label{L5}
\begin{split}
S_{\text{BEC}}=&\gamma_\beta\int dt\biggr[\sum\limits_{\mathbf{k}_\beta}\bigr|\dot{\psi}_{\mathbf{k}_\beta}(t)\bigr|^2-c_s^2\Bigr[\eta_{ij}+\frac{2\kappa}{\sqrt{V}}\sum\limits_{\mathbf{k},s}h_{\mathbf{k},s}(t)\\&\times\epsilon^s_{ij}(\mathbf{k})\Bigr]\sum\limits_{\mathbf{k}_\beta}k_\beta^ik_\beta^j\bigr|\psi_{\mathbf{k}_\beta}(t)\bigr|^2\biggr]
\end{split}
\end{equation}
where $\gamma_\beta$ has the dimension of length in natural units and $\gamma_\beta\equiv\frac{V_\beta}{2\lambda}(3\tilde{\sigma}^2-m^2)$. The total action of the theory can be obtained by combining eq.(\ref{L2}) with eq.(\ref{L5}).  Making use of the principle of least action ($\frac{\delta S}{\delta \psi^*_{\mathbf{k}_\beta}}=0$), we obtain the equation of motion for $\psi_{\mathbf{k}_\beta}(t)$ as
\begin{equation}\label{L6}
\begin{split}
\ddot{\psi}_{\mathbf{k}_\beta}(t)+c_s^2\Bigr[\eta_{ij}+\frac{2\kappa}{\sqrt{V}}\sum\limits_{\mathbf{k},s}h_{\mathbf{k},s}(t)\epsilon^s_{ij}(\mathbf{k})\Bigr]{k}^i_\beta {k}^j_\beta\psi_{\mathbf{k}_\beta}(t)&=0~.
\end{split}
\end{equation}
Again extremizing the total action with respect to the variable $h^*_{\mathbf{k},s}$, we obtain the equation of motion correponding to the gravitational fluctuation part as
\begin{equation}\label{L7}
\begin{split}
\ddot{h}_{\mathbf{k},s}(t)+k^2h_{\mathbf{k},s}(t)=-\frac{4\gamma_\beta\kappa c_s^2}{\sqrt{V}}\epsilon^{s*}_{ij}(\mathbf{k})\sum\limits_{\mathbf{k}_\beta} k_\beta^ik_\beta^j\left\lvert\psi_{\mathbf{k}_\beta}(t)\right\rvert^2.
\end{split}
\end{equation}
The next step is to raise the phase space variables to operator status and impose canonical commutation relation between the conjugate variables. In the interaction picture, the fluctuation of the gravitational quantum fluctuation takes the form \cite{KannoSodaTokuda}, $\delta \hat{h}^I_{\mathbf{k},s}(t)=\hat{h}_{\mathbf{k},s}^I(t)-h_{\text{cl}}^s(\mathbf{k},t).$ The quantum gravitational fluctuation term in terms of the creation and annihilation operators reads
\begin{equation}\label{L8}
\hat{h}_{\mathbf{k},s}^I(t)=u_k(t)\hat{a}_s(\mathbf{k})+u_k^*(t)\hat{a}^\dagger_s(-\mathbf{k})
\end{equation} 
where $k=|\mathbf{k}|$ and $u_k(t)$ is the mode function.  It is now possible to write down the graviton-noise induced quantum mechanical equation of motion corresponding to eq.(\ref{L6}) as
\begin{equation}\label{L9}
\begin{split}
&\ddot{\hat{\psi}}_{\mathbf{k}_\beta}(t)+c_s^2\left(\eta_{ij}+h_{ij}^{\text{cl}}(t,0)+\delta\hat{N}_{ij}(t)\right)k_\beta^i k_\beta^j\hat{\psi}_{\mathbf{k}_\beta}(t)=0
\end{split}
\end{equation}
where we have dropped all unnecessary contributions and we have made use of the following definitons
\begin{align}
h_{ij}^{\text{cl}}(t,\mathbf{x})\equiv&\frac{2\kappa}{\sqrt{V}}\sum\limits_{s}\smashoperator{\sum\limits_{\substack{{\mathbf{k}}\\{|\mathbf{k}|\leq\Omega_m}}}}h_{\text{cl}}^s(\mathbf{k},t)e^{i\mathbf{k}\cdot \mathbf{x}}\epsilon^s_{ij}(\mathbf{k})\label{L10}\\
\delta\hat{N}_{ij}(t)\equiv&\frac{2\kappa }{\sqrt{V}}\sum\limits_{s}\smashoperator{\sum\limits_{\substack{{\mathbf{k}}\\{|\mathbf{k}|\leq\Omega_m}}}}\delta \hat{h}^I_{\mathbf{k},s}(t)\epsilon^s_{ij}(\mathbf{k})~.\label{L11}
\end{align}
It is easy to infer from eq.(\ref{L9}) that the equation of motion is Langevin-like and has a stochastic nature due to the noise induced by gravitons. Solving eq.(\ref{L11}) and taking contributions from the noise fluctuation term at the final time of measurement only, one arrives at the following solution for the time-dependent part of the single-mode pseudo-Goldstone boson as
\begin{equation}\label{L12}
\begin{split}
\hat{\psi}_{\mathbf{k}_\beta}(t)=\hat{\alpha}^\beta(\tau)e^{-i\omega_\beta t}+\hat{\beta}^{\beta}(\tau)e^{i\omega_\beta t}
\end{split}
\end{equation}
where the modified coefficients $\hat{\alpha}^\beta(\tau)$ and $\hat{\beta}^\beta(\tau)$ are defined as
\begin{align}
&\hat{\alpha}^\beta(\tau)\equiv1-\frac{{k_\beta}_x^2-{k_\beta}_y^2}{4k_\beta^2}(1+2i\omega_\beta \tau)\delta\hat{N}(\tau)\label{L13}~,\\
&\hat{\beta}^\beta(\tau)\equiv\frac{{k_\beta}_x^2-{k_\beta}_y^2}{4k_\beta^2}\delta\hat{N}(\tau)\nonumber\\&+\frac{{k_\beta}_x^2-{k_\beta}_y^2}{4k_\beta^2}\sqrt{\pi}\varepsilon\omega_\beta\tau\left[e^{-\frac{\tau^2}{4}(\Omega-2\omega_\beta)^2}-e^{-\frac{\tau^2}{4}(\Omega+2\omega_\beta)^2}\right]\label{L14}
\end{align}
with $h^{\text{cl}}(t,0)=\varepsilon e^{-\frac{t^2}{\tau^2}}\sin\Omega t $. Our primary aim is to calculate the standard deviation in the measurement of the gravity wave amplitude parameter $\varepsilon$. The idea is to make use of quantum metrology techniques to obtain the quantum gravity modifed Fisher information corresponding to the same amplitude parameter.
\section{Quantum metrology and the noise of gravitons}\label{S4}
\begin{figure}[ht!]
\begin{center}
\includegraphics[scale=0.26]{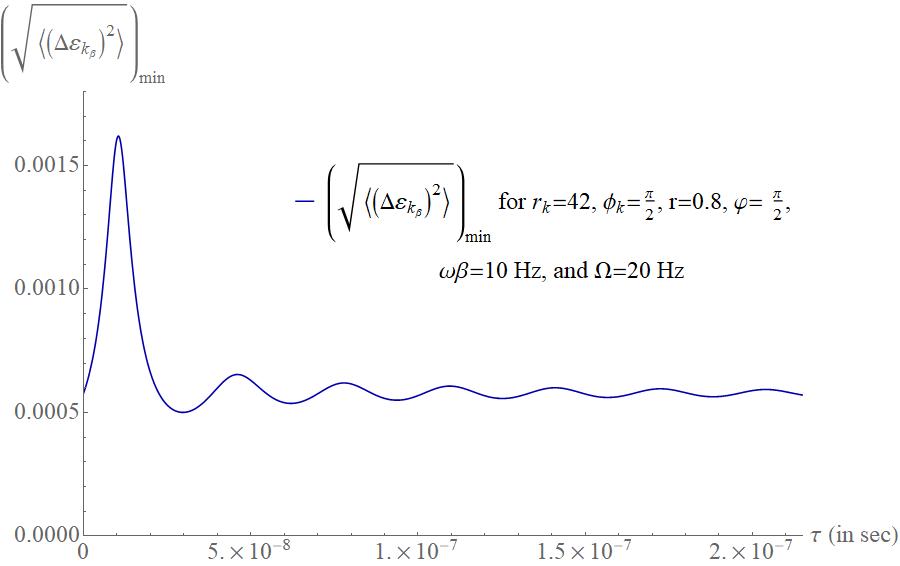}
\caption{$\sqrt{\langle(\Delta\varepsilon_{k_\beta})^2\rangle}_{\text{min.}}$ vs $\tau$ plot for the initial state of the graviton being in a highly squeezed state.\label{Fish1fr42}}
\end{center}
\end{figure}

\begin{figure}[ht!]
\begin{center}
\includegraphics[scale=0.242]{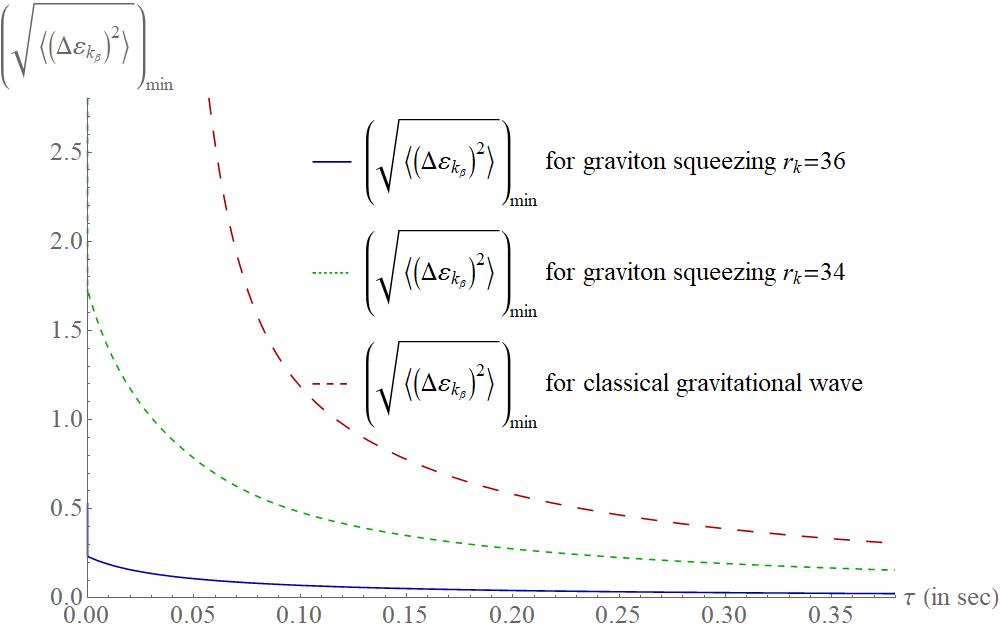}
\caption{$\sqrt{\langle(\Delta\varepsilon_{k_\beta})^2\rangle}_{\text{min.}}$ vs $\tau$ plot for squeezed gravitons with squeezing $r_k=34$ and $r_k=36$ respectively against the case of a classical gravitational wave.\label{Fish2frcomp}}
\end{center}
\end{figure}

\begin{figure}[ht!]
\begin{center}
\includegraphics[scale=0.242]{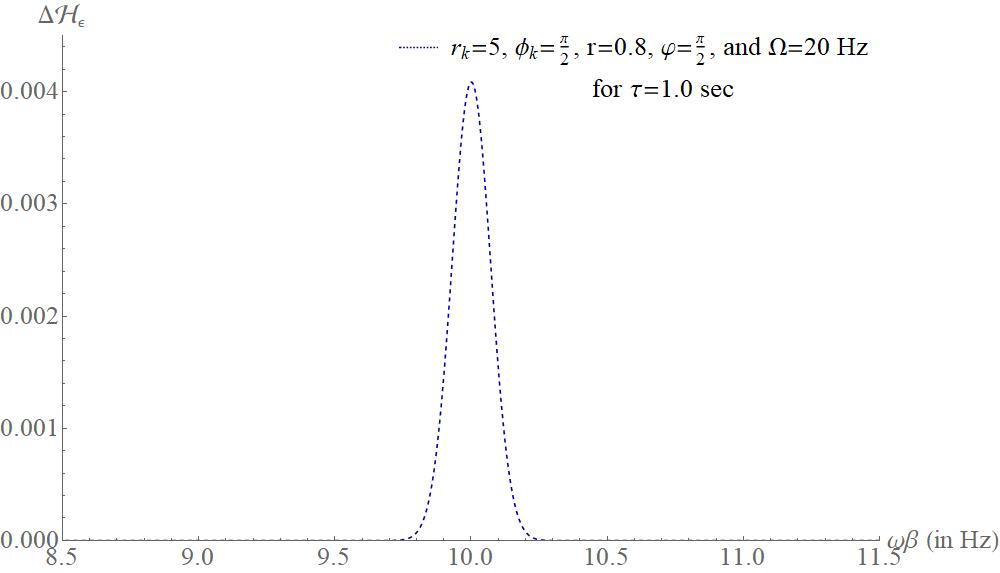}
\caption{$\Delta\mathcal{H}_\varepsilon$ vs $\omega_\beta$ plot for $\Omega_m=10^8$ Hz, $\Omega=100$ Hz, $r_k=5$, $\phi_k=\frac{\pi}{2}$, $r=0.8$, and $\varphi=\frac{\pi}{2}$. From the figure we can see that the peak of the standard deviation in the QGFI is observed near the resonance point $\omega_\beta=\frac{\Omega}{2}=10$ Hz.}\label{ResonanceO20}
\end{center}
\end{figure}

\begin{figure}[ht!]
\begin{center}
\includegraphics[scale=0.182]{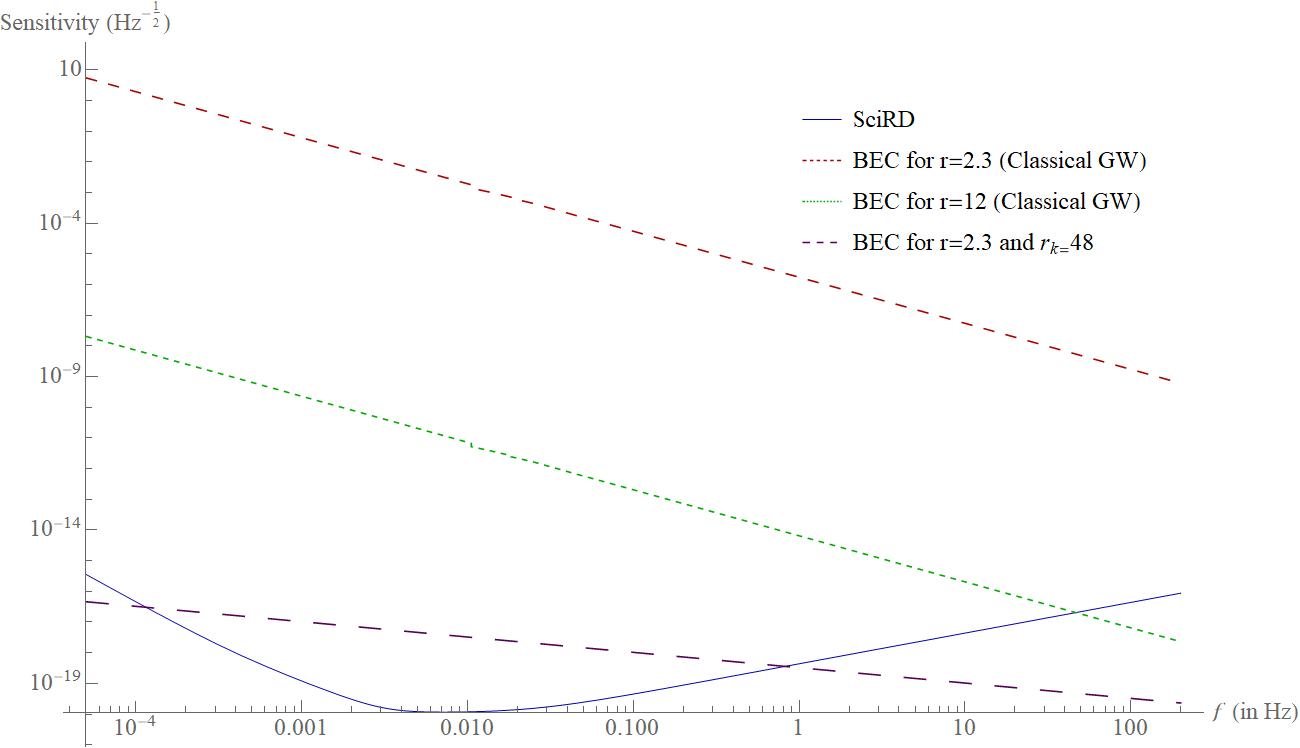}
\caption{The SciRD sensitivity formula is plotted along with  the BEC model sensitivity formula against the wave frequency $f$. We plot the case of BEC-Classical Gravity wave model alongside the BEC-Graviton model. \label{BEC_Graviton_Detector2}}
\end{center}
\end{figure}

\noindent It is easier to describe a Bose-Einstein condensate by a covariance matrix. For a single mode vacuum state of bosons in thermal equilibrium with a density matrix corresponding to the $k_\beta$-th mode of the system $\hat{\rho}=e^{-\text{\ss}\hat{H}}/\text{tr}[e^{-\text{\ss}\hat{H}}]$, the covariance matrix after phonon squeezing has been obtained and reads \cite{PhononBEC3, continuous_variable_QI} 
\begin{equation}\label{L15}
\begin{split}
&\Sigma_{\text{sq.}}[T]=\left(\frac{2\mathcal{N}+1}{2}\right)q\\&\times
\begin{bmatrix}
\cosh 2r+\cos\varphi\sinh 2r&&\sin\varphi\sinh 2r\\
\sin\varphi\sinh 2r&&\cosh 2r-\cos\varphi\sinh 2r
\end{bmatrix}
\end{split}
\end{equation} 
where $\mathcal{N}=\frac{1}{e^{\text{\ss}}-1}$ with $\text{\ss}=\frac{1}{k_BT}$ and the squeezing parameter is $r$ with squeezing angle $\varphi$. If the temperature now is set to zero in eq.(\ref{L15}), we arrive at the squeezed covariance matrix of the single mode Bose-Einstein condensate.
After the interaction with a gravitational wave, the covariance matrix corresponding to the single-mode BEC transforms as \cite{quantum_metrology}
\begin{equation}\label{L16}
\begin{split}
\tilde{\Sigma}_k(\tilde{\varepsilon})=\mathcal{M}_{kk}(\tilde{\varepsilon})\Sigma_{\text{sq.}}[0]\mathcal{M}_{kk}^T(\tilde{\varepsilon})+\sum\limits_{j\neq k}\mathcal{M}_{kj}(\tilde{\varepsilon})\mathcal{M}^T_{kj}(\tilde{\varepsilon})
\end{split}
\end{equation}
where $\tilde{\varepsilon}=\varepsilon\frac{{k_\beta}_x^2-{k_\beta}_y^2}{k_\beta^2}$ and the symplectic matrix $\mathcal{M}_{kj}(\tilde{\varepsilon})$ is given as \cite{PhononBEC3,quantum_metrology}
\begin{equation}\label{L17}
\mathcal{M}_{kj}(\tilde{\varepsilon})=\begin{bmatrix}
\Re[{\alpha}^\beta_{kj}-{\beta}^\beta_{kj}]&&\Im[{\alpha}^\beta_{kj}+{\beta}^\beta_{kj}]\\
-\Im[{\alpha}^\beta_{kj}-{\beta}^\beta_{kj}]&&\Re[{\alpha}^\beta_{kj}+{\beta}^\beta_{kj}]
\end{bmatrix}
\end{equation}
with ${\alpha}^\beta_{kj}$ and $\beta^\beta_{kj}$ denoting the classical  Bogoliubov coefficients. It is important to note that because of the Bogoliubov coefficients being modified by the quantum gravity correction, the matrix $\mathcal{M}$ will change as well leading to an overall graviton induced noise effect in the covariance matrix of the squeezed BEC. The Fisher information in terms of the elements of the covariance matrix takes a simple form given as (with no displacement of the squeezed states) \cite{Uhlmann2mode,quantum_metrology}
\begin{equation}\label{L18}
\begin{split}
\mathcal{H}_\varepsilon=4\mathcal{C}^{(2)}=&2\left(\Sigma_{11}^{(0)}\Sigma_{22}^{(2)}+\Sigma_{11}^{(2)}\Sigma_{22}^{(0)}-2\Sigma_{12}^{(0)}\Sigma_{12}^{(2)}\right)\\&+\frac{1}{2}\left(\Sigma_{11}^{(1)}\Sigma_{22}^{(1)}-2\Sigma_{12}^{(1)}\Sigma_{12}^{(1)}\right)~.
\end{split}
\end{equation}
In our current analysis, because of the noise fluctuation terms involved, the quantum Fisher information  cannot be measured directly but one needs to measure the stochastic average of the same with respect to some initial graviton state. We term this modified Fisher information as quantum gravitational Fisher information (QGFI) and the stochastic average of the same reads
\begin{equation}\label{L19}
\begin{split}
\llangle \hat{\mathcal{H}}_{\varepsilon}\rrangle=\mathcal{H}_\varepsilon^{(0)}+\frac{\llangle\{\delta\hat{N}(\tau),\delta\hat{N}(\tau)\}\rrangle}{32\varepsilon^2}\mathcal{H}_\varepsilon^{(2)}
\end{split}
\end{equation}
where $\mathcal{H}_\varepsilon^{(0)}$ is the quantum Fisher information when the gravitational wave contribution is treated classically \cite{PhononBEC3}. The squeezing angle of the phonons can be set to $\varphi=\frac{\pi}{2}$ which is possible to achieve experimentally \cite{Chelkowski,Johnsson}. For the graviton initially being in a squeezed state, the stochastic average of QGFI takes the form
\begin{equation}\label{L20}
\begin{split}
&\llangle\hat{\mathcal{H}}_\varepsilon\rrangle=\mathcal{H}_\varepsilon^{(0)}+\frac{l_p^2\Omega_m^2}{15 \pi\varepsilon^2c^2}\mathcal{B}(r_k,\phi_k,\tau)\mathcal{H}_{\varepsilon}^{(2)}\\
&=\frac{1}{64}\pi\omega_\beta^2\tau^2\left(
e^{2\omega_\beta\Omega\tau^2}-1\right)^2
e^{-\frac{\tau^2}{2}(\Omega+2\omega_\beta)^2}(1+\cosh4r\\&+4\sinh^22r)+\frac{l_p^2\Omega_m^2}{30 \pi\varepsilon^2c^2}\bigr(3+2\omega_\beta^2\tau^2+\cosh 4r+6\omega_\beta\tau\\&\times\sinh4r
+6\omega_\beta^2\tau^2\cosh4r\bigr)\biggr(\cosh 2r_k
+\frac{1}{2\Omega_m^2\tau^2}\sinh 2r_k\\&\left(\cos\phi_k-\cos(2\Omega_m\tau-\phi_k)-2\Omega_m\tau
\sin(2\Omega_m\tau-\phi_k)\right)\biggr)~.
\end{split}
\end{equation}
Eq.(\ref{L20}) is one of the main results in our paper and is given in the manuscript for the sake of completeness of the analysis. The inequality involving the minimum value of the standard deviation in the gravitational amplitude parameter reads
\begin{equation}\label{L21}
\langle (\Delta\varepsilon_{k_\beta})^2\rangle\geq\frac{15}{2\pi\mathfrak{N}\llangle\hat{\mathcal{H}}_\varepsilon\rrangle}
\end{equation}
with $\mathfrak{N}$ denoting the number of measurement of the gravitational wave using the Bose-Einstein condensate. From the above uncertainty we get,  the minimum value of $\sqrt{\langle(\Delta\varepsilon_{k_\beta})^2\rangle}$ as $\sqrt{\langle(\Delta\varepsilon_{k_\beta})^2\rangle}_{\text{min.}}=\sqrt{\frac{15}{2\pi\llangle\hat{\mathcal{H}}_\varepsilon\rrangle}}$. We consider a gravitational wave with frequency $20$ Hz with initial graviton squeezing $r_k=42$ and squeezing angle $\phi_k=\frac{\pi}{2}$ interacting with a Bose-Einstein condensate having a phonon squeezing $r=0.8$ with a squeezing angle $\varphi=\frac{\pi}{2}$. In order to detect a signature of the gravity wave, it is essential to have the minimum value of the uncertainty in the amplitude parameter below unity. The plot of the minimum value of the standard deviation in $\varepsilon_{k_\beta}$ vs the observation time $\tau$ is given is Fig.(\ref{Fish1fr42}). 
There are very important observations that one can deduce from Fig.(\ref{Fish1fr42}). The first important point to observe is that corresponding to a single-mode of the BEC even initially there is a possibility of detecting traces of the gravitational wave. A simple calculation yields that for a classical gravitational wave $\sqrt{\langle (\Delta\varepsilon_{k_\beta})^2\rangle}_{\text{min.}}\sim 2.91\times10^{17}$ whereas for a quantum gravitational scenario, the minimum value of the standard deviation in the amplitude parameter is of the order of $5.69\times 10^{-4}$ when $\tau=1\times 10^{-7}$ sec. For the minimum standard deviation of the amplitude to be less than unity, the measurement time needs to be of the order of 1 sec for the classical gravity wave-BEC interaction. In case of the quantum gravity scenario with high enough squeezing from the primordial gravitational waves \cite{KannoSodaTokuda,QGravNoise,QGravLett,QGravD}, it is always possible to detect signatures of a fluctuating gravitational background even at the very start of an experiment. Another important aspect is that for a linearized quantum gravity theory the graviton states can superpose with each other allowing a overall gravitational field background which is present from the beginning even when the gravity wave just appears as a result of the quantumness of the linearized gravity theory. This indicates that the BEC will be able to detect short time gravitational disturbance generated purely due to the existence of gravitons. This experimental detection must occur at a nano-micro second time period just after the commencement of the experiment using a BEC system which will indicate the existence of gravitons but the main problem is that such a measurement is very difficult and in order to do so, one needs to make multiple measurements and check for such short time initial measrements. Another important aspect is that the standard deviation in $\varepsilon_{k_\beta}$ corresponding to a single mode of the BEC is not infinite also initially, indicating a quantum nature of gravity. The amplitude of the standard deviation rapidly increases with a decay in the squeezing of the gravitons. One can also focus on increasing the squeezing of the phonons as well to increase the sensitivity of the measurement. It is although very important to note that with a decrease in the squeezing of the graviton state, the minimum value in the measurement of the standard deviation of the gravitational wave amplitude becomes very high indicating a non detectability of such a scenario. For a initial vacuum graviton state ($r_k=0$), we obtain $\lim\limits_{\tau\rightarrow0}\llangle\hat{\mathcal{H}}_\varepsilon\rrangle=\frac{l_p^2\Omega_m^2}{15 \pi\varepsilon^2c^2}\sim10^{-31}$ ($\Omega_m\sim10^8$ and $\varepsilon\sim 10^{-21}$). This indicates, $\sqrt{\langle(\Delta \varepsilon)^2\rangle}\geq \frac{15\varepsilon c}{l_p \Omega_m \sqrt{2}}\sim 10^{16}$. Such a high minimum value of the standard deviation in $\varepsilon_{k_\beta}$ parameter indicates a negative detection outcome.  For a better understanding of the squeezing dependence of such BEC involved detection scenario, we have plotted the minimum value of the standard deviation in amplitude parameter with respect to $\tau$ for $r_k=34,~36$ and a classical gravity wave in Fig.(\ref{Fish2frcomp}). We observe from Fig.(\ref{Fish2frcomp}) that with higher squeezing the uncertainty in the measurement of the amplitude parameter becomes lower leading to a higher chance of detection. 
The above result indicates that in the case a quantum gravity scenario, the minimum value of the standard deviation of the gravity wave amplitude parameter never becomes zero and can be arbitrarily reduced with squeezed graviton state indicating a higher chance at detecting quantum gravity signatures. One can also calculate the square of the standard deviation in the QGFI which can be expressed in an extended form given as ($\varphi=\frac{\pi}{2}$)
\begin{equation}\label{L22}
\begin{split}
&(\Delta\mathcal{H}_\varepsilon)^2=\frac{l_p^2\omega_\beta^2\Omega_m^2\tau^2}{960\varepsilon^2c^2}\bigr(e^{-\frac{\tau^2}{4}(\Omega-2\omega_\beta)^2}-e^{-\frac{\tau^2}{4}(\Omega+2\omega_\beta)^2}\bigr)^2\\&\left(2\cosh^22r+4\sinh^2 2r+6\omega_\beta \tau\sinh 4r\right)^2\mathcal{B}(r_k,\phi_k,\tau)
\end{split}
\end{equation}
with the form of $\mathcal{B}(r_k,\phi_k,\tau)$ as
\begin{equation}\label{L23}
\begin{split}
&\mathcal{B}(r_k,\phi_k,\tau)=\cosh 2r_k
+\frac{1}{2\Omega_m^2\tau^2}\sinh 2r_k\bigr(\cos\phi_k\\&-\cos(2\Omega_m\tau-\phi_k)-2\Omega_m\tau\sin(2\Omega_m\tau-\phi_k)\bigr)~.
\end{split}
\end{equation}
We shall now look at the behaviour of the standard deviation in the QGFI around resonance. For a finite measurement of $\tau=1$ sec, with $r_k=5,\phi_k=\frac{\pi}{2},r=0.8$, and $\varphi=\frac{\pi}{2}$, we plot $\Delta \mathcal{H}_\varepsilon$ against the phonon frequency $\omega_\beta$ for an incoming gravitational wave with frequency 20 Hz in Fig. (\ref{ResonanceO20}). The cut-off frequency for a primordial gravitational wave lies around $\Omega\sim10^8$ Hz.
We find out from Fig.(\ref{ResonanceO20}) that the standard deviation in the quantum gravitational Fisher information becomes maximum for the phonon frequency $\omega_\beta=\frac{\Omega}{2}=10$ Hz. It can also be observed by plotting (not presented here) that the standard deviation in the QGFI becomes maximum for a longer observation time. This is an expected feature as can be seen from eq.(\ref{L21}), the change in the variance of the amplitude parameter is inversely proportional to the square root of the stochastic average of the QGFI. As again can be observed from Fig.(\ref{Fish2frcomp}), with increasing time, the minimum value of the standard deviation in the amplitude parameter $\varepsilon_{k_\beta}$ gets significantly smaller, as a result the measurement in the amplitude parameter becomes more accurate. This results in the increase in the standard deviation of the QGFI. In order to depict a different scenario, the graviton noise is meddled in by a Gaussian time decay as $\delta\hat{N}(t')=\cos\Omega t' e^{-\frac{{t'}^2}{\tau^2}}\delta\hat{N}(t). $ Instead of restricting one at the single mode measurement scenario, one can sum over all possible phonon mode frequencies. In order to do so one needs to express $\omega_\beta=c_sk_\beta=\frac{\pi c_sn_\beta}{L_\beta}$ and integrate over all possible phonon numbers along with the first ocatant of the solid angle. A part of the integral diverges, which can be managed by setting the squeezing of the phonons to $r=\frac{1}{4}\cosh^{-1}(7)\simeq 0.66$. This makes the divergent part of the integral to vanish. Squeezing control is achievable experimentally and squeezing as high as $r\simeq 0.83$ ($7.2$ dB \cite{GuLiWuYang}) has already been attained experimentally. Phonon squeezing is generally expressed in terms of decibels. Now the position squeezing $s$ is related to the dimensionless squeezing parameter $r$ via the relation $s=-10\log_{10}[e^{-2r}]$\cite{Lvovsky}. Techniques like second order Raman scattering \cite{Raman1,Raman2} or pump-probe detection scheme \cite{Pump_Probe_Detection} can be used to squeeze cold bosonic atoms in optical lattices \cite{Cold_Bosonic_Optical_Lattice} (in the presence of an optical potential). One can now obtain the final form of the total standard deviation in the amplitude parameter  as
\begin{equation}\label{L24}
\begin{split}
&\frac{1}{\langle(\Delta\varepsilon)^2\rangle}\leq\frac{\mathfrak{N}V_\beta \mathfrak{r}_1}{7680\sqrt{2\pi}c_s^3\tau^3}(\Omega^4\tau^4+6\Omega^2\tau^2+3-3e^{-\frac{\Omega^2\tau^2}{2}})\\&+\frac{\hbar G\mathfrak{N}V_\beta \Omega_m^2\mathfrak{r}_1\mathcal{B}(\tau)}{225 \pi\sqrt{2\pi}\varepsilon^2c_s^3\tau^3 c^5}(\Omega^4\tau^4+6\Omega^2\tau^2+3+3e^{-\frac{\Omega^2\tau^2}{2}})
\end{split}
\end{equation}
where we have used $\mathcal{B}(\tau)\equiv \mathcal{B}(r_k,\phi_k,\tau)$ and $\mathfrak{r}_1=1+\cosh 4r+4\sinh^22r$.  
 Now it has been possible to create a BEC with length $L_\beta\sim 10^{-3}$ m \cite{Vengalattore,Barr,Greytak} and as a result $\tau=\frac{c}{L_\beta}\sim 10^{-11}$ sec. Even for $\Omega\sim\Omega_m$,  $\Omega\tau\sim 10^{-3}$ which indicates that $\Omega\tau\ll 1$ $\forall\Omega$. For a total observation time of $\tau_{\text{obs.}}$, one can run approximately $\mathfrak{N}\sim\frac{\tau_{\text{obs.}}}{\tau}$ number of observations. In this scenario, one can obtain a lower bound to the minimum value of the observation time for a vacuum graviton state without any squeezing and $\Omega\sim\Omega_m$ as 
\begin{equation}\label{L25}
\begin{split}
\tau_{\text{min.}}^0\simeq& \sqrt{\frac{2}{3\pi}}\frac{32l_p\Omega_m}{c\varepsilon\Omega}\Bigr\rvert_{\Omega\rightarrow\Omega_m}\simeq 1.59\times 10^{-22}\text{ sec}.
\end{split}
\end{equation}
This is one of the most important results in our paper. It was argued in \cite{PhononBEC3} that the observation time cannot be arbitrarily smaller by making a sensitivity analysis of the BEC setup with available experimental data from gravitational wave detection. Because of the induced quantum gravitational corrections in the theory, we obtain a theoretical lower bound to the minimum observaton time which indicates that below $\tau^0_{\text{min.}}$, it is not possible to detect any signature of gravitational fluctuation. It is again very important to note that the Gaussian decay term in the classical gravity wave part  appears from the template of the gravitational wave whereas for the current analysis it is an imposed condition. One can now compare the sensitivity of the BEC-graviton model with the projected sensitivitiy mode of the upcoming LISA observatory (Laser Interferometer Space Antenna). We have here used the SciRD sensitivity model for comparing with the BEC detector sensitivity \cite{SciRD,SciRD2}. The sensitivity corresponding to the BEC-graviton model is governed by $\sqrt{\langle (\Delta\varepsilon)^2\rangle}/\sqrt{f}~\text{Hz}^{-\frac{1}{2}}$ for an incoming gravitational wave with frequency $f$ Hz with the resonance frequency for the phonons being $f/2$ Hz. In Fig.(\ref{BEC_Graviton_Detector2}), we have plotted the SciRD sensitivity ($\sqrt{S_{\text{h,SciRD}}(f)}~\text{Hz}^{-\frac{1}{2}}$) with the Bose-Einstein condensate models for the classical gravitational wave case as well as graviton case. From Fig.(\ref{BEC_Graviton_Detector2}), we observe that a BEC with phonon squeezing $r=2.3$, the BEC is not at all sensitive for classical gravitational waves at lower frequency regimes ($10^{-4}-100$ Hz). It is although very important to note that when the gravitons are considered into the picture with squeezing as high as $48$, the BEC sensitivity is in line with that of the projected sensitivity of the LISA observatory. Hence, for a future generation of gravitational wave detectors, if a primordial gravitational wave is detected then a detection by a BEC with low phonon squeezing will confirm the signature of graviton. 
For the final part of our analysis, we consider interaction between individual phonon modes in the BEC. This leads to a decoherence in the system\footnote{For an introduction see \cite{Serafinietal} and see \cite{SBECOTM} for a detailed and extended analysis of this letter.}. For a bosonic system at low temperature, Beliaev damping is significant and in the zero temperature limit takes the form \cite{Beliaev} $\gamma\simeq \frac{3}{640\pi}\frac{\hbar \omega_\beta^5}{m_\beta n_\beta c_s^5}$, where $n_\beta$ denotes the number density of the atoms in the BEC and $m_\beta$ denotes the mass of each individual atoms. It was observed that the speed of sound is  $c_s\simeq1.2\times10^{-2} \text{ m sec}^{-1}$ in a BEC with number  density $7\times 10^{20}\text{ m}^{-3}$\cite{Andrewsetal}. 
\begin{figure}
\begin{center}
\includegraphics[scale=0.244]{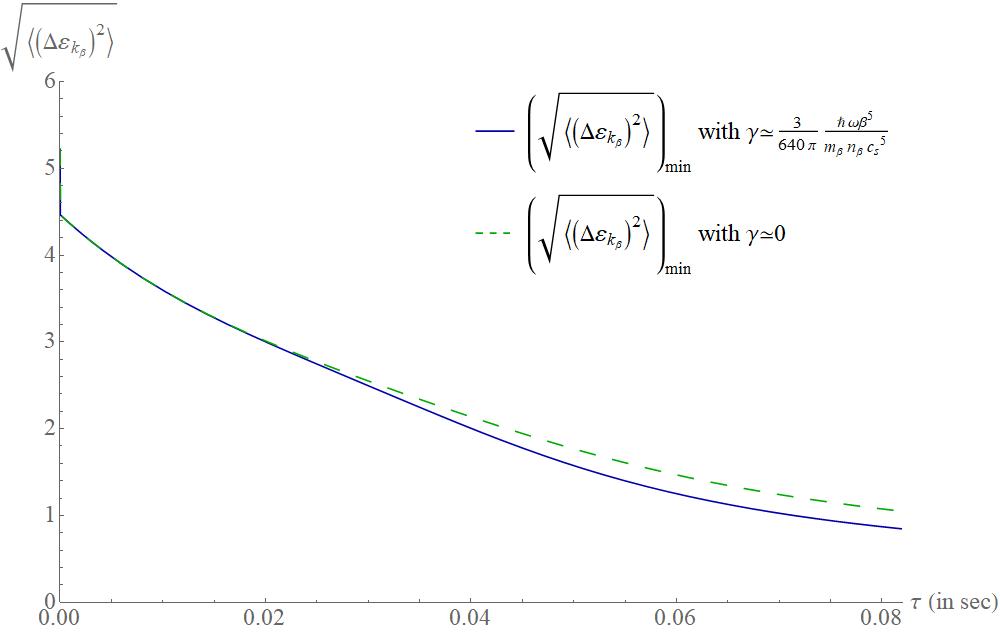}
\caption{$\sqrt{\langle(\Delta \varepsilon_{k_\beta})^2\rangle}_{\text{min.}}$ vs $\tau$ has been plotted for the case with and without damping with $r_k=33$.\label{Beliaevp1}}
\end{center}
\end{figure}
\noindent The value of the Beliaev damping factor for the phonon mode frequency $\omega_\beta= 10$ Hz reads $\gamma\simeq 9.034\times 10^{-19} \text{ sec}^{-1}$. The BEC will then mainly be sensitive to $\Omega=20$ Hz gravitational waves. We then plot the minimum value of the standard deviation in the amplitude parameter against the observation time $\tau$ for the squeezing of the graviton $r_k=33$.  We can see that the plot with a non-vanishing decoherence term differs from the usual non-damping case at around $\tau=0.02$ sec. For a very small squeezing with $r_k=2$, separation happens at very early times which is of the order of  $\tau\sim10^{-7}$ sec. For a complete and rigorous description of decoherence due to interacting phonon modes, one needs to comply to the prescription executed in \cite{HowlFuentes}. Another very important consequence of such an experiment is that one needs to keep on producing BEC at a continuous rate. Such experiments can be done using magneto-optical traps, the experimental setup of which has been proposed in \cite{Ketterle} and later observed in \cite{Tiecke}. In this work, we are thoroughly looking for signatures of quantum gravity rather than direct evidence of the quantum nature of gravity. The continuation of this paper will evolve towards finding a conclusive evidence of the quantum nature of gravity if preliminary signatures of quantum gravity are found  experimentally.
\section{Conclusion}\label{S7}
\noindent In this letter, we consider a Bose-Einstein condensate interacting with a gravitational wave where the gravitational fluctuation is quantized incorporating quantum gravitational effects into the theory. Integrating the heavy fields out of the theory, and quantizing the theory by means of raising the phase space variables to operator status and imposing canonical commutation relation between the conjugate variables, we arrive at a Bose-Einstein condensate model integrating with a quantum gravitational fluctuation. Using the principle of least action, we obtain an equation of motion corresponding to the time-dependent part of the pseudo-Goldstone bosons and an equation of motion correponding to the graviton mode. Because of the noise induced by the gravitons the equation of motion becomes Langevin-like. Solving the equation of motion corresponding to the bosonic part, we obtain a solution for the time dependent part of the pseudo-Goldstone boson which in principle gives the form of the pseudo-Goldstone boson as a whole. It is important to note that because of the stochastic parameters involved the Bogoliubov coefficients also become dependent on the stochastic noise parameter. Finally, we obtain the quantum gravity modified Fisher information. Because of its stochastic nature, we term it as the quantum gravitational Fisher information (QGFI). The stochastic average of the QGFI with respect to the initial graviton state is an observable. Using the quantum mechanical extension of the Cram\'{e}r-Rao bound we obtain an uncertainty involving the minimum value of the standard deviation in the gravitational amplitude parameter and the stochastic average of the QGFI. Next, we observe that the standard deviation for the amplitude parameter becomes very small even at initial times which is a very high quantity when classical gravitational waves are being considered. If after a nano-micro second order single measurement time a gravitational fluctuation signal is detected via means of the resonance condition, it is always possible to claim it as a signature of quantum gravity. The important aspect is that one needs to make multiple measurements of short time periods to truly confirm the signature of such a quantum gravitational effect via the use of a single mode Bose-Einstein condensate. We further inspect the standard deviation in the measurement of the Fisher information which indicates that with the increase in the single measurement time, the standard deviation becomes substantially larger at the resonance point  which may be possible to detect as well. Then we have considered a noise template infused by a Gaussian decay factor and integrated over all possible mode values. From the positivity condition of the minimum value of the standard deviation in the amplitude parameter, we have obtained a minimum measurement time which is of the order of $10^{-22}$ sec. Below this time, one will not be able to detect signature of a gravitational fluctuation. Finally, we consider decoherence effects into the theory due to interacting phonon modes. We observe that the minimum value of the standard deviation shows a delay in splitting from the non-damping scenario  for a higher squeezing factor. In this work, we mainly focus on the signatures of the quantum gravity when the phonons of the Bose-Einstem condensate are in resonance with the gravitational fluctuation. In the next work,  \textit{``Zweite Abhandlung"}, we shall be more focussed on the quantum nature of gravity from the perspective of an experimental standpoint and will propose a novel experiment to probe for conclusive evidence by detecting the existence of a linearized quantum gravity theory.

\end{document}